\documentclass[apj]{emulateapj}

\submitted{ApJ, accepted 17/07/2009}

\usepackage{amsmath}
\usepackage{color}

\newcommand{\be}{\begin{equation}}
\newcommand{\ee}{\end{equation}}
\newcommand{\bea}{\begin{eqnarray}}
\newcommand{\eea}{\end{eqnarray}}
\newcommand{\bes}{\be\begin{split}}
\newcommand{\eesp}{\end{split}\ee}
\newcommand{\ha}{HI}
\newcommand{\hm}{H$_2$}

\newcommand{\ket}[1]{|#1\rangle}

\newcommand{\h}{h}
\newcommand{\freq}{\nu}
\newcommand{\fobs}{\freq_{\rm o}}
\newcommand{\frest}{\freq_{\rm e}}

\newcommand{\fco}{\freq_{\rm CO}}

\newcommand{\qhm}{q_{\rm H_2}}

\newcommand{\hp}{h_{\rm p}}
\newcommand{\kb}{k_{\rm b}}

\newcommand{\f}{R_{\rm mol}}
\newcommand{\fc}{\f^{\rm c}}

\newcommand{\rd}{r_{\rm disk}}

\newcommand{\rhm}{r_{\rm H_2}}

\newcommand{\rhmhalfmass}{\rhm^{\rm half}}

\newcommand{\mass}{M}
\newcommand{\msun}{{\rm M}_{\odot}}
\newcommand{\mhydro}{\mass_{\rm H}}

\newcommand{\mg}{\mass_{\rm gas}}

\newcommand{\mhm}{\mass_{{\rm H}_2}}

\newcommand{\masshm}{m_{{\rm H}_2}}
\newcommand{\nhm}{N_{{\rm H}_2}}

\newcommand{\mz}{\mass_{\rm Z}}

\newcommand{\Sigmah}{\Sigma_{\rm H}}
\newcommand{\Sigmaha}{\Sigma_{\rm HI}}
\newcommand{\Sigmahm}{\Sigma_{\rm H_2}}

\newcommand{\Sigmastd}{\tilde{\Sigma}_{\rm H}}

\newcommand{\lsun}{{\rm L}_{\odot}}

\newcommand{\dl}{D_{\rm L}}
\newcommand{\da}{D_{\rm A}}
\newcommand{\lfreq}{L}
\newcommand{\lvel}{L^{\rm V}}
\newcommand{\lbt}{L^{\rm T}}
\newcommand{\sfreq}{S}
\newcommand{\svel}{S^{\rm V}}

\newcommand{\tb}{T_{\rm B}}
\newcommand{\wlrest}{\lambda_{\rm e}}
\newcommand{\wlobs}{\lambda_{\rm o}}
\newcommand{\vel}{V}

\newcommand{\velint}{W}
\newcommand{\x}{X}
\newcommand{\xj}{\x_{\rm J}}
\newcommand{\alphaj}{\alpha_{\rm J}}
\newcommand{\ibt}{I}
\newcommand{\tcmb}{T_{\rm CMB}}
\newcommand{\tsb}{T_{\rm SB}}
\newcommand{\tsbmax}{\tsb^{\rm max}}
\newcommand{\sigsf}{\Sigma_{\rm SF}}
\newcommand{\sigsfc}{\sigsf^{\rm c}}
\newcommand{\tagn}{T_{\rm AGN}}
\newcommand{\tagnmax}{\tagn^{\rm max}}
\newcommand{\mdotbh}{\dot{M}_{\rm BH}}
\newcommand{\mdotbhc}{\mdotbh^{\rm c}}
\newcommand{\fclumpy}{f_{\rm clumpy}}
\newcommand{\overlap}{B}
\newcommand{\filling}{F}

\newcommand{\partition}{\mathcal{Z}}
\newcommand{\te}{T_{\rm e}}

\newcommand{\opt}{\tau}
\newcommand{\optj}{\opt_{\rm J}}
\newcommand{\optc}{\opt_{\rm c}}

\newcommand{\feff}{\varepsilon}
\newcommand{\transj}{J\!\rightarrow\!J\!-\!1}
\newcommand{\cov}{\kappa}

\newcommand{\sigcrit}{\Sigma_{\rm c}}

\shorttitle{CO-luminosity functions}
\shortauthors{Obreschkow et al.}

\begin{document}

\title{Prediction of the Cosmic Evolution of the CO-Luminosity Functions}

\author{D. Obreschkow, I. Heywood, H.-R. Kl\"ockner, and S. Rawlings}
\affil{Astrophysics, Department of Physics, University of Oxford, Keble Road, Oxford, OX1 3RH, UK}

\begin{abstract}
We predict the emission line luminosity functions (LFs) of the first 10 rotational transitions of $\rm ^{12}C^{16}O$ in galaxies at redshift $z=0$ to $z=10$. This prediction relies on a recently presented simulation of the molecular cold gas content in $\sim3\cdot10^7$ evolving galaxies based on the Millennium Simulation. We combine this simulation with a model for the conversion between molecular mass and CO-line intensities, which incorporates the following mechanisms: (i) molecular gas is heated by the CMB, starbursts (SBs), and active galactic nuclei (AGNs); (ii) molecular clouds in dense or inclined galaxies can overlap; (iii) compact gas can attain a smooth distribution in the densest part of disks; (iv) CO-luminosities scale with metallicity changes between galaxies; (v) CO-luminosities are always detected against the CMB. We analyze the relative importance of these effects and predict the cosmic evolution of the CO-LFs. The most notable conclusion is that the detection of regular galaxies (i.e.~no AGN, no massive SB) at high $z\gtrsim7$ in CO-emission will be dramatically hindered by the weak contrast against the CMB, in contradiction to earlier claims that CMB-heating will ease the detection of high-redshift CO. The full simulation of extragalactic CO-lines and the predicted CO-LFs at any redshift can be accessed online\footnote{http://s-cubed.physics.ox.ac.uk/, go to ``S$^3$-SAX-Box''} and they should be useful for the modeling of CO-line surveys with future telescopes, such as ALMA, the LMT, and the SKA.
\end{abstract}

\keywords{galaxies: high-redshift --- galaxies: evolution --- ISM: atoms --- ISM: molecules}

\maketitle

\section{Introduction}\label{section_introduction}

An increasing body of evidence suggests that molecular hydrogen (\hm) widely dominated over atomic hydrogen (\ha) in the regular galaxies of the early universe \citep[e.g.][]{Obreschkow2009c}. Empirical corner stones towards this conclusion were the measurement of strong CO-line emission in distant regular galaxies \citep{Daddi2008}, the detection of \ha~via Lyman-$\alpha$ absorption against distant quasars \citep[e.g.][]{Prochaska2005}, the observational confirmation of a correlation between the interstellar gas pressure and \hm/\ha-ratios \citep{Blitz2006}, and the observational confirmation that galaxy sizes increase significantly with cosmic time \citep[e.g.][]{Bouwens2004}.

In light of future millimeter/submillimeter telescopes, such as the Atacama Large Millimeter/submillimeter Array (ALMA) and the Large Millimeter Telescope (LMT), much attention is directed towards the possibility of detecting the suspected molecular gas (mostly \hm) at high redshift via the characteristic emission lines of the CO-molecule. However, the case for frequent CO-detections in regular high-redshift galaxies is by no means secure, since neither the cosmic evolution of the \hm-mass function (MF), nor the evolution of the relationship between \hm-masses and CO-line luminosities is well constrained to-date. An elucidation of this situation seems nevertheless within reach, owing to a long list of specific discoveries over the past two decades (Section \ref{section_method}), based on which computer simulations could already predict the CO-line emission of individual high-redshift galaxies in some detail \citep[e.g.][]{Combes1999,Greve2008}. Moreover, \cite{Blain2000} and \cite{Carilli2002} predicted the number of detectable CO-sources in various frequency ranges. They assumed that the CO-line luminosities evolve with the far-infrared (FIR) luminosity, and they tackled the cosmic evolution of the FIR-luminosity function (LF) by considering a pure density evolution. While this approach is perhaps justified at low redshifts, it probably oversimplifies the physical complexity of CO-emission at high redshift ($z>1$) as we shall show in this paper.

A missing jigsaw piece in the bigger picture is a physical prediction of the cosmic evolution of the galaxy LFs for different CO-emission lines. In this paper, we will attempt such a prediction by concatenating many specific empirical and theoretical findings about \hm~and CO. The two main steps towards our prediction of the CO-LFs are (i) a model for the cosmic evolution of the \hm-MF and (ii) a model for the conversion between \hm-masses and CO-line luminosities. In this paper, we shall focus on the latter, while adopting the \hm-masses of a sample of $\sim3\cdot10^7$ galaxies \citep{Obreschkow2009b}, simulated based on the Millennium dark matter simulation \citep{Springel2005}.

In Section \ref{section_simulation}, we summarize the galaxy simulation producing the \hm-masses and various other galaxy-properties related to CO-line emission. Our model for CO-line luminosities is developed in Section \ref{section_method}. Section \ref{section_results} presents the prediction of the cosmic evolution of the CO-LFs and discusses their dependence on the mechanisms listed at the beginning of Section \ref{section_method}. Section \ref{section_discussion} ranks the relative importance of these mechanisms and discusses the limitations of their implementation. A brief summary is given in Section \ref{section_conclusion}.

\section{Simulation of the \hm-MF}\label{section_simulation}

This section summarizes the cold gas simulation presented in \cite{Obreschkow2009b}. Main results and limitations were discussed in detail by \cite{Obreschkow2009b} and \cite{Obreschkow2009c}.

The simulation has three consecutive layers. The first layer is the Millennium Simulation \citep{Springel2005}, an $N$-body dark matter simulation in a periodic box of comoving volume $(500\,\h^{-1}\,{\rm Mpc})^3$, where $H_0=100\,\h\rm\,km\,s^{-1}\,Mpc^{-1}$ and $\h=0.73$. The second simulation layer uses the evolving mass skeleton of the Millennium Simulation to tackle the formation and cosmic evolution of galaxies in a semi-analytic fashion \citep{Croton2006,DeLucia2007}. This is a global approach, where galaxies are represented by a list of global properties, such as position, velocity, and total masses of gas, stars, and black holes. These properties were evolved using simplistic formulae for mechanisms, such as gas cooling, reionization, star formation, gas heating by supernovae, starbursts (SBs), black hole accretion, black hole coalescence, and the formation of stellar bulges via disk instabilities. The resulting virtual galaxy catalog (hereafter the ``DeLucia-catalog'') contains the positions, velocities, merger histories, and intrinsic properties of $\sim3\cdot10^7$ galaxies at 64 cosmic time steps. At redshift $z=0$, galaxies as low in mass as the Small Magellanic Cloud are resolved. The free parameters in the semi-analytic model were tuned to various observations in the local universe (see \citealp{Croton2006}). Therefore, despite the simplistic implementation and the possible incompleteness of this model, the simulated galaxies nonetheless provide a good fit to the joint luminosity/color/morphology distribution of observed low-redshift galaxies \citep{Cole2001,Huang2003,Norberg2002}, the bulge-to-black hole mass relation \citep{Haering2004}, the Tully--Fisher relation \citep{Giovanelli1997}, and the cold gas metallicity as a function of stellar mass \citep{Tremonti2004}.

The cold gas masses of the simulated galaxies are the net result of gas accretion by cooling from a hot halo (dominant mode) and galaxy mergers, gas losses by star formation and feedback from supernovae, and cooling flow suppression by feedback from accreting black holes. The DeLucia-catalog does not distinguish between molecular and atomic cold gas, but simplistically treats all cold gas as a single phase. Therefore, the third simulation layer, explained by \cite{Obreschkow2009b}, consists of post-processing the DeLucia-catalog to split the cold gas masses of each galaxy into \ha, \hm, and He. Our model for this subdivision mainly relies on three empirical findings: (i) Most cold gas in regular spiral \citep{Leroy2008} and elliptical galaxies \citep{Young2002} in the local universe resides in flat disks, and there is evidence that this feature extends to higher redshifts \citep[e.g.][]{Tacconi2006}. (ii) The surface density of the total hydrogen component (\ha+\hm) is approximately described by an axially symmetric exponential profile \citep{Leroy2008},
\be
  \Sigmah(r) = \Sigmastd\,\exp(-r/\rd)\,, \label{eqsigmah}
\ee
where $\rd$ is the exponential scale length and the normalization factor $\Sigmastd$ can be calculated as $\Sigmastd\equiv\mhydro/(2\pi\rd^2)$, where $\mhydro$ is the total mass of cold hydrogen in the disk. (iii) The local \hm/\ha-mass ratio closely follows the gas pressure of the interstellar medium outside molecular clouds over at least four orders of magnitude in pressure and for various galaxy types \citep{Blitz2006,Leroy2008}. Based on those findings, we \citep{Obreschkow2009b} derived an analytic expression of the \ha- and \hm-surface density profiles,
\bea
  \Sigmaha(r) & = & \frac{\Sigmastd\,\exp(-r/\rd)}{1+\fc\exp(-1.6\,r/\rd)}\,, \label{eqsigmaha} \\
  \Sigmahm(r) & = & \frac{\Sigmastd\,\fc\,\exp(-2.6\,r/\rd)}{1+\fc\exp(-1.6\,r/\rd)}\,, \label{eqsigmahm}
\eea
where $\fc$ is the \hm/\ha-mass ratio at the galaxy center. This model was applied to the galaxies in the DeLucia-catalog to characterize their \ha~and \hm~content (masses, diameters, and circular velocities). The resulting hydrogen simulation successfully reproduces many local observations of \ha~and \hm, such as MFs, mass--diameter relations, and mass--velocity relations \citep{Obreschkow2009b}. This success is quite surprising, since our model for \ha~and \hm~only introduced one additional free parameter to match the observed average space density of cold gas in the local universe \citep{Obreschkow2009b}. A key prediction of this simulation is that the \hm/\ha-ratio of most regular galaxies increases dramatically with redshift, hence causing a clear signature of cosmic ``downsizing'' in the \hm-MF \citep{Obreschkow2009c}, i.e.~a negative shift in the mass scale with cosmic time.

The simulated \hm-MF at $z=0$ approximately matches the local \hm-MF inferred from the local CO(1--0)-LF \citep{Keres2003,Obreschkow2009a}, and the few measurements of CO-line emission from regular galaxies at $z\approx1.5$ \citep{Daddi2008} are consistent with the predicted \hm-MF at this redshift \citep{Obreschkow2009c}. Furthermore, the predicted comoving space density of \hm~evolves proportionally to the observed space density of star formation rates \citep[e.g.][]{Hopkins2007} within a factor 2 out to at least $z=3$. For those reasons, we expect the simulated \hm-MF to scale reasonably well with redshift. Yet, at $z\gtrsim5$ the simulation becomes very uncertain because the geometries and matter content of regular galaxies are virtually unconstrained from an empirical viewpoint. The young age and short merger intervals of these galaxies compared to their dynamical time scales, may have caused them to deviate substantially from the simplistic disk-gas model. An extended discussion of these and other limitations at low and high redshift is given in Section 6.3 of \cite{Obreschkow2009b}.

\section{Model for the CO/\hm~conversion}\label{section_method}

Most detections of \hm~rely on emission lines originating from the relaxation of the rotational $J$-levels of the $\rm ^{12}C^{16}O$-molecule (hereafter ``CO''). Appendix \ref{appendix_backgroundx} provides background information on the inference of \hm-masses from CO-line measurements and highlights the justification and drawbacks of this method.

To predict the CO-line luminosities associated with the molecular gas masses of the simulated galaxies (Section \ref{section_simulation}), we shall now introduce a simplistic but physically motivated model for the conversion between \hm-masses and CO-luminosities at any redshift. This model aims to respect the following theoretical and empirical constraints:
\begin{enumerate}
  \item The temperature of molecular gas depends on the temperature of the cosmic microwave background (CMB) and on the radiative feedback from SBs and active galactic nuclei (AGNs).
  \item Molecular clumps can shield each other if they overlap along the line of sight and in velocity space. This effect may not be negligible in the dense galaxies at high redshift, especially if observed edge-on.
  \item While locally observed molecular gas is organized in giant molecular clouds (GMCs), the dense gas in compact luminous galaxies, such as ultra luminous infrared galaxies (ULIRGs), is predicted to follow a smooth distribution.
  \item The CO-line emission of molecular gas is correlated with the CO/\hm-mass ratio, i.e.~to the metallicity of the galaxy.
  \item The CMB presents an observing background. The absorption of CO-lines against the CMB may significantly reduce the effectively detectable luminosities of CO-emission lines.
\end{enumerate}
These mechanisms will be modeled one by one over Sections \ref{subsection_cosed}--\ref{subsection_cmb_absorption}.

\subsection{Gas temperature and the CO-ladder}\label{subsection_cosed}

\begin{figure}[h]
  \includegraphics[width=0.98\columnwidth]{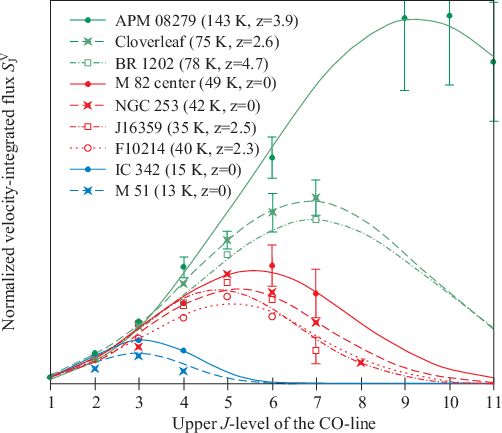}
  \caption{CO-SEDs of nine nearby and distant galaxies: APM~08279+5255
 \citep{Weiss2007}, Cloverleaf H1413+117 \citep{Barvainis1997}, BR~1202-0725 \citep{Omont1996,Kawabe1999}, the central region of M\,82 \citep{Weiss2005b}, NGC\,253 \citep{Guesten2006}, SMM~J16359+6612 \citep{Weiss2005a}, IRAS F10214+4724 \citep{Ao2008}, IC\,342 \citep{Israel2003}, M\,51 \citep{Wielebinski1999,Nieten1999}. Symbols and error bars represent the measurements. To avoid confusion, only some error bars are displayed. Different colors are used for regular galaxies (blue), SBs (red), and QSOs (green). Lines represent fits of Eq.~(\ref{eqmodel1}) with $\optc=2$ and excitation temperatures $\te$ shown in the legend. Note that the CO-SEDs use different vertical scales, \textit{not} exactly normalized to the respective CO(1--0) fluxes. The vertical scale of the models has been optimized simultaneously with $\te$ to fit all data points, including CO(1--0), within their uncertainties.}
  \label{fig_co_sed}
\end{figure}

To model the luminosity-ratios of different CO-lines, we analyzed the CO-spectral energy distributions (SEDs) of nine galaxies drawn from the literature (see Fig.~\ref{fig_co_sed} and references therein). This sample includes local regular galaxies, local and distant SBs, and distant quasi stellar objects (QSOs). Four of these sources (SMM~J16359+6612, F10214+4724, APM~08279+5255, Cloverleaf H1413+135) are known to be strongly magnified by gravitational lensing. We assume that this has no major effect on the flux-ratios between different CO-lines. This assumption relies on the fact that the lensed galaxies are FIR-bright objects, which makes it likely that the strongly lensed regions include the star-bursting ones. Those are also the high-excitation (HE) regions, which seem to dominate the CO-emission of most CO-lines (see discussion of M\,82 in this section).

Surprisingly, all nine CO-SEDs are well fitted by a model for a single gas component in local thermodynamic equilibrium (LTE). In Appendix \ref{appendix_thermal_model}, we show that the frequency-integrated line luminosities (= power) of such a model scales with the upper level $J$ of the transition as
\be\label{eqmodel1}
  L_{\rm J} \propto \left[1-\exp(-\optj)\right]\cdot\frac{J^4}{\exp\left(\frac{\hp\,\fco\,J}{\kb\,\te}\right)-1},
\ee
where $\te$ is the excitation temperature, $\fco=115\rm\,GHz$ is the rest-frame frequency of the CO(1--0)-transition, and $\optj$ is the optical depth. The latter scales with $J$ as
\be\label{eqmodel2}
  \optj = 7.2\,\optc\,\exp\!\left(\!-\frac{\hp\,\fco\,J^2}{2\,\kb\,\te}\right)\sinh\!\left(\frac{\hp\,\fco\,J}{2\,\kb\,\te}\right),
\ee
where $\optc$ is a constant. The factor 7.2 in Eq.~(\ref{eqmodel2}) was introduced in order for $\optc$ to correspond to the optical depth of the CO(1--0) line (i.e.~$\opt_1=\optc$) at the excitation temperature $\te\approx17\rm\,K$, which is the lowest temperature of our model (see end of this section).

If normalized to the CO(1--0)-luminosity $L_1$, the LTE-model of Eqs.~(\ref{eqmodel1}, \ref{eqmodel2}) has two free parameters $\optc$ and $\te$. In order to apply this model to the velocity-integrated fluxes $\svel_{\rm J}$ shown in Fig.~\ref{fig_co_sed}, we use the relation $L_{\rm J}\propto J\,\svel_{\rm J}$ (see Appendix \ref{appendix_luminosities}, Eq.~\ref{eqpresolomon1}).

We first fitted the LTE-model to the observed CO-SEDs individually via $\chi^2$-minimization. The resulting 1-$\sigma$ confidence intervals of the temperatures $\te$ equal $10-20\%$ of their best-fit values. Despite this uncertainty, a clear dependence of $\te$ on the galaxy types (regular, SB, QSO) can de detected (see below). By contrast, the parameters $\optc$ are poorly constrained. Their best-fit values range from $0.5$ to $5$ with no clear trend amongst the different galaxy types, and their confidence intervals are such that a single parameter $\optc$ for all CO-SEDs seems to provide a consistent solution. We therefore tested a second model, where all nine SEDs share the same parameter $\optc$, and found that the Bayesian evidence \citep[e.g.][]{Sivia2006} of this 10-parameter model ($1\times\optc$, $9\times\te$) against the 18-parameter model ($9\times\optc$, $9\times\te$) is ``strong'' with odds of order $10^5:1$. We therefore assume a single parameter $\optc$ for all galaxies, emphasizing, however, that the actual optical depth $\optj$ varies considerably as a function of $J$ and $\te$ by virtue of Eq.~(\ref{eqmodel2}). The best fit to all nine SEDs yields $\optc=2$, consistent with the moderate optical depths for different CO-lines found by \cite{Barvainis1997} in the Cloverleaf quasar. The excitation temperatures $\te$ corresponding to $\optc=2$ are listed in Fig.~\ref{fig_co_sed}. The individual reduced $\chi^2$'s for each galaxy range from $0.5$ to $1.2$, hence demonstrating that the LTE-model with a single parameter $\optc$ provides an excellent fit to all observed CO-line-ratios.

This conclusion justifies the use of the single component LTE-model as a \emph{working model}, but it does not imply that this model describes the \emph{physical reality} of molecular gas. In fact, it seems that neither the assumption of LTE-conditions, nor that of a single gas component are fully satisfied in reality.

First, the density of molecular gas is often too low to collisionally excite the higher rotational levels to thermal equilibrium. In this case, the LTE-conditions are not met. A more accurate description of the excitation state is then provided by the so-called large velocity gradient (LVG) models \citep{DeJong1975}, which are more complex than the LTE-model. A plausible explanation for the surprising success of the LTE-model is that the suppression of high-$J$ emission by sub-thermal excitation can be approximately mimicked by a thermalized gas with a slightly underestimated optical depth, or a temperature $\te$ slightly below the kinetic temperature of the gas. Another explanation is that in real clouds the sub-thermal excitation of high-$J$ states could be compensated by a minor fraction of much warmer ($\sim\!100\rm\,K$) and denser molecular gas, such as is seen next to the star-forming cores in nearby molecular clouds \citep[e.g.~the ``ridge'' in the Orion molecular cloud,][]{Lis2003}. In any case, we stress that the temperatures $\te$ identified by our LTE-analysis should not be considered as very accurate. Better models, albeit more complex, can be found in the references of Fig.~\ref{fig_co_sed}.

Second, the assumption of a single component seems to work because in most galaxies one component widely dominates the total CO-SED. A good example to illustrate this conclusion and its limitations is the nearby SB M\,82, for which the CO-SED up to the CO(7--6)-transition has been presented by \cite{Weiss2005b}. The CO-SED of the center of M\,82 is displayed in Fig.~\ref{fig_co_sed} and is reasonably well described by a LTE-model (reduced $\chi^2=1.1$). Yet, an in-depth LVG-analysis \citep[Fig.~7 in][]{Weiss2005b} revealed that the center of M\,82 exhibits a low-excitation (LE) and an HE component with kinetic temperatures of $\sim50\rm\,K$ (perhaps higher for the LE component), consistent with the single temperature of the LTE-model of $49\rm\,K$ (see Fig.~\ref{fig_co_sed}). In terms of velocity-integrated fluxes, the CO-SED of the LE component peaks around the upper level $J=3-4$, while the HE component peaks around $J=6-7$. However, the flux from the HE component completely dominates the combined CO-SED, such that the latter still peaks around $J=6-7$. This domination of the HE component justifies the use of a single component as a working model. Moreover, the domination of the HE component, in which the excitation conditions are close to LTE, adds another reason for the aforementioned success of the LTE-model. The limitations of the single component model become obvious, when considering the CO-SED of the entire galaxy M\,82 \citep{Weiss2005b}. The exceptionally strong gas outflows from the star-bursting center add an additional LE component, which dominates the total CO-SED up to the CO(3--2)-transition or perhaps the CO(4--3)-transition. The success of the single component model for the other three SBs in Fig.~\ref{fig_co_sed} suggests that the strongly CO-luminous outflows of M\,82 are rather anomalous.

Despite the above limitations of the single component LTE-model, we shall use this model for the rest of this paper for three reasons: (i) given current computational resources, applying an LVG-model to $\sim10^9$ galaxies (i.e.~up to $\sim3\cdot10^7$ galaxies per discrete time step) is highly impractical; (ii) as demonstrated above (e.g.~Fig.~\ref{fig_co_sed}), the LTE-model is a reasonable working model in the sense that it can approximately fit most observed CO-SEDs; and (iii) the differences between the LTE-model and the LVG-model are often much smaller than the uncertainties associated with other mechanisms, such as cloud overlap in high-$z$ galaxies, metallicity, or gas heating by SBs (see~Section \ref{subsection_limitations}).

Fig.~\ref{fig_co_sed} demonstrates that the characteristic excitation temperatures $\te$ increase from regular galaxies to SBs, and more so to QSOs. This supports the interpretation of gas-heating by radiation from SBs and AGNs (see also observations of \citealp{Weiss2007} and theoretical work of \citealp{Maloney1988}). On the other hand, $\te$ must also depend on the temperature of the CMB at the redshift of the source \citep{Silk1997,Combes1999}.

We assume that in regular galaxies the molecular gas is heated by a constant specific power (i.e.~power per unit gas mass), representing the intra-cloud radiative heating by massive stars and supernovae associated with regular star formation efficiencies. This specific power implies a minimal temperature $T_0$ for the bulk of the molecular gas. In addition, the CMB represents a background temperature of $\tcmb(z)=(1+z)\cdot2.7\rm\,K$. If the radiative heating of molecular gas happens via absorption by optically thick dust, then the resulting gas temperature or the CO-excitation temperature is $\te^4\approx T_0^4+\tcmb(z)^4$, as can be seen from combining the Stefan-Boltzmann law with the conservation of energy. Following the same argument, we can also include the heating of SBs and AGNs via
\be\label{eqexcitationt}
  \te^4 = T_0^4+\tcmb^4(z)+\tsb^4+\tagn^4,
\ee
where $\tsb$ and $\tagn$ are galaxy-dependent parameters characterizing the estimated temperatures of the molecular gas, if respectively SB-feedback or AGN-feedback were the only sources of radiative heating. \cite{Combes1999} pointed out that, if the radiative transfer is mediated by optically thin dust with an optical depth proportional to $\lambda^{-2}$, the exponents in Eq.~(\ref{eqexcitationt}) should be increased from 4 to 6. In reality the exponents in Eq.~(\ref{eqexcitationt}) are therefore likely to be somewhat higher than 4. Yet, Eq.~(\ref{eqexcitationt}) only depends on the precise value of the exponents in the few cases where the highest temperatures on the right-hand-side are comparable, while otherwise the highest temperature completely dominates $\te$.

In the following, we require that the specific radiation power ($\propto\tsb^4$) acquired by the molecular gas from SBs, increases proportionally to the surface density of the star formation rate (SFR) $\sigsf$ for small values of $\sigsf$, while saturating at an upper limit, characterized by the temperature $\tsbmax$. This saturation level encodes all possible self-regulation mechanisms, preventing further heating, such as the suppression of star formation by photo-dissociation of molecular gas. To parameterize the efficiency of SB-heating, we define the characteristic SFR-density $\sigsfc$, at which the specific radiation power reaches $50\%$ of the saturation level. A minimal parametrization of these requirements is given by the function
\be\label{eqparatsb}
  \tsb^4 = {\tsbmax}^4\,\sigsf/(\sigsf+\sigsfc),
\ee
which reduces to the linear relation $\tsb^4\approx{\tsbmax}^4\,\sigsf/\sigsfc$ for $\sigsf\ll\sigsfc$. To compute $\sigsf={\rm SFR}/(\pi\,r_{\rm SF}^2)$ for the galaxies in our simulation, we approximate the characteristic length $r_{\rm SF}$ with the half-mass radius $\rhmhalfmass$ of molecular gas and we use the SFRs computed by the semi-analytic model \citep[see][]{Croton2006}. In this model stars can form via two mechanisms: (i) quiescent continual star formation in the disk, which depends on the cold gas surface density; (ii) star-bursting activity in the bulge, which is driven by galaxy mergers. We shall use the combined SFRs of both modes to calculate $\sigsf$, since, in principle, both modes are likely to cause inter-cloud radiative heating, if the corresponding SFR densities are high enough, i.e.~of order $\sigsfc$.

In analogy to SBs, we parameterize the heating from AGNs via
\be\label{eqparatagn}
  \tagn^4 = {\tagnmax}^4\,\mdotbh/(\mdotbh+\mdotbhc),
\ee
where $\tagnmax$ is the maximal CO-excitation temperature that can be achieved by AGN-heating, $\mdotbh$ is the black hole mass accretion rate, and $\mdotbhc$ is the critical accretion rate, where the specific heating power is half the maximum value. In the semi-analytic model of the DeLucia-catalog \citep{Croton2006}, black holes can grow via two mechanisms: (i) a quiescent mode, whereby black holes continually accrete material from a static hot halo; (ii) a merger mode, where the black holes of merging galaxies coalesce, while accreting additional material from the cold gas disks. The free parameters in this model, were adjusted such that the predicted relation between black hole mass and bulge mass matches the local observations by \cite{Haering2004}. Since our model for CO-heating only depends on $\mdotbh$, we have implicitly assumed that all growing black holes have the same heating efficiency, independent of their growth mode and physical parameters, such as the black hole mass -- a simplistic assumption, which may well require a more careful treatment as large samples of CO-detected AGN become available.


To finalize our model, we need to estimate the five parameters $T_0$, $\tsbmax$, $\tagnmax$, $\sigsfc$, and $\mdotbhc$. To fix $T_0$, we consider regular galaxies (no SB, no AGN) in the local universe ($\tcmb(z=0)=2.7\,K$), where Eq.~(\ref{eqexcitationt}) implies that $\te$ is nearly identical to $T_0$. From simultaneous CO(2--1) and CO(1--0) detections in 35 regular galaxies in the local universe, \citet{Braine1993b} concluded that the ratio between the brightness temperature luminosities is $\lbt_2/\lbt_1=0.89$ with a scatter of only 0.06. According to Eq.~(\ref{eq4}), this is equivalent to $L_2/L_1=2^3\cdot0.89$, which, by virtue of Eqs.~(\ref{eqmodel1}, \ref{eqmodel2}), implies a one-to-one correspondence between $\te\approx T_0$ and $\optc$. If we impose $\optc=2$ (see above), then $T_0\approx17\rm\,K$, which roughly agrees with the excitation temperatures of the regular galaxies M\,51 and IC\,342 for the same depth parameter (see Fig.~\ref{fig_co_sed}). We therefore fix $T_0\equiv17\rm\,K$.

We further set the critical star-formation density to $\sigsfc\equiv500\,\msun\rm\,yr^{-1}\,kpc^{-2}$, consistent with observations of the nuclear SBs of M\,82 \citep{deGrijs2001} and NGC\,253 \citep{Beck1984}. For those galaxies Eq.~(\ref{eqparatsb}) then implies that ${\tsbmax}^4=\tsb^4/2$, where $\tsb\approx\te=40-50\rm\,K$ (see Fig.~\ref{fig_co_sed}), hence $\tsbmax=50-60\rm\,K$. We therefore choose $\tsbmax\equiv60\,K$. Our chosen value for $\sigsfc$ also compares well to the star-formation density $\sim10^3\,\msun\rm\,yr^{-1}\,kpc^{-2}$ predicted by \cite{Thompson2005} for the optically thick, dense regions of star forming disks.

For AGN heating, we choose $\tagnmax\equiv150\rm\,K$, assuming that the QSO APM~08279+5255 represents an object close to the maximal possible heating. The critical black hole accretion rate $\mdotbhc$ is assumed to be $\mdotbhc\equiv10\,\msun\rm\,yr^{-1}$, consistent with the higher Eddington accretion rates in the sample of 121 radio-loud quasars studied by \citet{Bao2008}. Assuming a standard radiative accretion efficiency of $10\%$, this value for $\mdotbhc$ corresponds to a black hole mass of $5\cdot10^8\,\msun$, which is on the order of a typical progenitor of the supermassive black holes found in the massive galaxies in the local universe.

\subsection{Overlap of molecular clumps}\label{subsection_overlap}

A reason, why CO-radiation can be used as a \emph{linear} tracer of molecular mass in nearby galaxies despite its optical thickness is that most lines-of-sight to the molecular clumps in nearby galaxies do not cross other clumps, and hence CO behaves as if it were optically thin (see Appendix \ref{appendix_backgroundx}). However, at high redshift, galaxies are denser \citep[e.g.][]{Bouwens2004} and carry more molecular gas \citep{Obreschkow2009c}, and thus the overlap (in space and velocity) of molecular clumps may become significant. Such overlap will (i) reduce the directly visible surface area per unit molecular mass, and (ii) increase the effective optical depth of the CO-radiation.

\citeauthor{Bally1987} (\citeyear{Bally1987}; see also \citealp{Genzel1989}) identified and analyzed more than 100 clumps in the Orion molecular cloud. Based on these data, we assume that the diameters and masses of clumps are approximately $r_{\rm clump}=1\rm\,pc$ and $\mass_{\rm clump}=500\,\msun$. Given a total molecular mass $\mhm$, the number of clumps is approximated by
\be\label{eqmhmn}
  N = \frac{\mhm}{\mass_{\rm clump}}
\ee
and the ``fractional filling factor'' of a single clump can be approximated as
\be
  \cov = \frac{0.1\,r_{\rm clump}^2}{\qhm\,\rhm^2},
\ee
where $\qhm$ is the axes-ratio of the inclined galaxy-disk\footnote{The simulated DeLucia-catalog does not provide galaxy orientations. We therefore assign inclinations randomly between $0\rm\,deg$ (face-on) and $90\rm\,deg$ (edge-on) according to a sine-distribution.} as seen by the observer and $\rhm$ is the radius of molecular gas in this disk, which we take as the galactocentric radius, where $\Sigmahm(r)$ given in Eq.~(\ref{eqsigmahm}) equals $10\%$ of the maximal surface density. The factor 0.1 accounts for the fact that clouds are only considered to ``overlap'', if they have comparable radial velocities, as otherwise they become mutually transparent to CO-line radiation. The value of 0.1 is a rough estimate based on the velocity dispersion of each clump ($1-10\rm\,km\,s^{-1}$, \citealp{Bally1987} and \citealp{Maddalena1986}) and the fact that the relative velocities between two clumps can vary from $10\rm\,km\,s^{-1}$ (if in the same cloud-complex) up to the circular velocity of the disk of several $100\rm\,km\,s^{-1}$ (if in different parts of the galaxy). In a more accurate model, the value 0.1 would have to be altered with the inclination of the galaxy.

Assuming that the clumps are randomly distributed in space and frequency, we find that the fractional volume of the position--velocity space covered by the $N$ clumps, counting overlapping regions only once, is
\be\label{eqfilling}
  \filling=1-(1-\cov)^{\rm N}.
\ee
$\filling\in[0,1]$ is here called the ``filling factor'' (although other definitions of this term exist) and its expression of Eq.~(\ref{eqfilling}) can be derived iteratively by realizing that the filling factor of $i$ clumps, $i>1$, is $\cov_{\rm i}=\cov_{\rm i-1}+\cov\cdot(1-\cov_{\rm i-1})$ with $\cov_1\equiv\cov$. Since the summed volume occupied by all clumps in the position--velocity space equals $N\,\cov$, we find that any line-of-sight crossing at least one clump, must on average cross
\be
  \overlap = \frac{N\,\cov}{1-(1-\cov)^{\rm N}}.
\ee
clumps, which also overlap in velocity space. If clumps do not overlap (i.e.~$\overlap=1$), the emergent CO-line luminosities are proportional to the number of clumps $N$, and hence proportional to the molecular mass $\mhm$ (see Eq.~\ref{eqmhmn}). However, if the clumps overlap (i.e.~$\overlap>1$), the directly visible surface area of the molecular gas is proportional to $\mhm/\overlap$ and the optical depth increases from $\optj$ to $\overlap\,\optj$.

\subsection{Clumpy and smooth molecular gas}\label{subsection_densegas}

Measurements of CO-emission lines in distant ULIRGs revealed that the use of CO-luminosity-to-\hm-mass conversion factors known from local galaxies leads to \hm-masses on the order of or larger than the dynamical masses inferred from the circular velocities \citep{Scoville1991,Solomon2005}. This contradiction and high-resolution CO-maps of ULIRGs led to the new understanding that the densely packed GMCs at the center of massive compact galaxies are unstable against the tidal shear and therefore disintegrate into a smooth blend of gas and stars. Detailed observations and geometrical models of \cite{Downes1993} and \cite{Downes1998} uncovered that the smooth gas is about 5-times more CO-luminous per unit molecular mass. Multiple line observations of the two nearby ULIRGs Arp 220 and NGC 6240 \citep[][and references therein]{Greve2009} seem to confirm this model, but they also demonstrate that molecular gas in the smooth phase may coexist with clumpy gas (i.e.~with GMCs) in the outskirts of the nuclear disk.

To account for the possibility of smooth molecular gas, we assume that CO-luminosities per unit molecular mass scale proportionally to the efficiency
\be\label{eqmeff}
  \feff = \fclumpy+5\cdot(1-\fclumpy),
\ee
where $\fclumpy$ is the \hm-mass fraction in the regular clumpy phase (i.e.~in GMCs) and $(1-\fclumpy)$ is the \hm-mass fraction in the regular phase. We define the transition between the clumpy and the smooth gas phase at the \hm-surface density threshold $\sigcrit=10^3\,\msun\rm\,pc^{-2}$, which is between the highest \hm-densities observed in the local universe ($\sim10^2\,\msun\rm\,pc^{-2}$, e.g.~NGC 6946, \citealp{Leroy2008}) and the most extreme \hm-surface densities of ULIRGs ($\sim10^4\,\msun\rm\,pc^{-2}$, e.g.~Arp 220 and NGC 6240, \citealp{Greve2009}). Assuming a thickness of the nuclear disk of a few $10\rm\,pc$, consistent with the nuclear disk model of \cite{Downes1998}, the adopted value of value of $\sigcrit$ corresponds to a volume density of $\gtrsim10^3\rm\,cm^{-3}$. This value falls in between the two volume densities found by \cite{Greve2009} for the smooth and dense gas phases of Arp 220 and NGC 6240.

In the smooth phase, the \hm/\ha-mass ratio is much larger than unity, so that the \hm-surface density $\Sigmahm(r)$ can be safely approximated by the total hydrogen density $\Sigmah(r)$ given in Eq.~(\ref{eqsigmah}). The \hm-mass fraction in regions less dense than $\sigcrit$, can then be calculated as
\be
  \fclumpy = \left\{\begin{array}{ll}
  \frac{\sigcrit}{\Sigmastd}\left[1+\ln\left(\frac{\Sigmastd}{\sigcrit}\right)\right] & {\rm if~}\Sigmastd>\sigcrit, \\
  1 & \rm otherwise.
  \end{array}\right.
\ee

\begin{figure}[h]
  \includegraphics[width=\columnwidth]{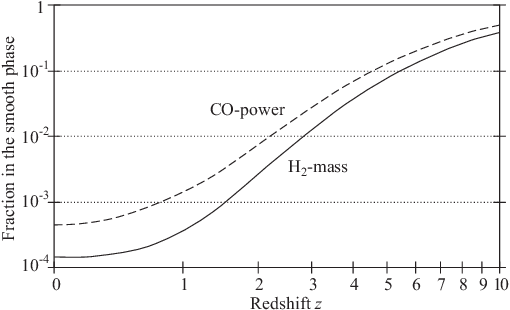}
  \caption{Cosmic evolution of the \hm-mass fraction in the smooth gas phase summed over all galaxies in the simulation (solid line), and the corresponding fraction of CO-line power (dashed line).}
  \label{fig_fdense_evolution}
\end{figure}

We shall assume that the overlap factor $\overlap$ is calculated in the same way for the smooth component as for the clumpy one, which corresponds to approximating the self-shielding of the smooth region by the self-shielding of a densely packed distribution of clumps with the same total volume and mass\footnote{The filling factor $\filling$ of the smooth component turns out to be very close to 1, and hence $\overlap\approx N\,\cov$.}. Fig.~\ref{fig_fdense_evolution} shows the simulated global fraction of \hm-mass in the smooth phase and the fraction of CO-power from this phase as a function of cosmic time. The predicted monotonic increase of both fractions with redshift clearly reflects the strong density and size evolution of cold gas disks predicted by the simulation \citep[e.g.][]{Obreschkow2009d}.

At $z=2$, the \hm-mass fraction in the smooth phase is about $0.3\%$ (corresponding to a fractional CO-power of $\sim1\%$), roughly consistent with the fact that the space density of ULIRGs is $\sim1\%$ of the space density of normal galaxies at this redshift \citep{Daddi2008}. The remaining $99\%$ of CO-power at $z=2$ in the simulation stems from clumpy gas, i.e.~from GMCs. This result seems consistent with recent observational evidence that star formation in many active star forming galaxies at $z\approx2$ is distributed on significantly larger scales than in ULIRGs \citep{Daddi2008,Genzel2008} and that star formation properties in high-$z$ galaxies are similar to those in GMCs \citep[e.g.][]{Gao2008}.

At $z=5$, the CO-power from the smooth gas phase is predicted to make up $\sim10\%$ of the total CO-power of all galaxies. One might argue that such small fractions can be neglected. However, in Section \ref{subsection_effects_smooth}, we will show that the contribution of CO-radiation from smooth molecular gas at $z=5$ will change the space density of the brightest objects in the CO-LFs by an order of magnitude.

\subsection{Metallicity}\label{subsection_metallicity}

Various recent observations revealed significant variations of the CO-luminosity-to-\hm-mass~conversion factor within and amongst the Milky Way (MW) and several nearby galaxies \citep{Arimoto1996,Boselli2002,Wilson1995,Israel2000,Paglione2001}. In general, the CO-luminosity per unit molecular mass turned out to be roughly proportional to the mass-fraction of metals \citep{Obreschkow2009a}. Such a dependence may naively be expected for radiation emitted by a metallic\footnote{Here, all elements other than hydrogen and helium are referred to as ``metals''.} molecule like CO, but considering the optical thickness of this radiation, one could also conclude that the CO-luminosity per unit molecular mass is nearly independent of the metallicity \citep{Kutner1985}. However, lower metallicities imply a lower dust-to-gas ratio and hence a more efficient destruction of CO by ultra violet (UV) radiation, which can restore a positive correlation between metallicities and the CO-luminosities \citep[see][and references therein]{Maloney1988}.

Based on these quantitative measurements and qualitative theoretical considerations, we decided to scale the luminosity of all CO-lines proportionally to $Z\equiv\mz/\mg$, where $\mz$ is the metal mass in cold gas and $\mg$ is the total cold gas mass (including He).

\subsection{Effective luminosity against the CMB}\label{subsection_cmb_absorption}

The CMB power per unit frequency has its maximum within the frequency band covered by the CO-lines. For example, at $z=0$ the CMB peaks between the CO(1--0)-line and the CO(2--1)-line, and at $z=5$ the CMB peaks between the CO(8--7)-line and the CO(9--8)-line. Therefore, the absorption of CO-lines against the CMB may significantly reduce the detectable luminosities of CO-emission lines.

Within our assumption that clouds are in local thermal equilibrium, Kirchhoff's law of thermal radiation globally predicts that the absorptivity of the clouds equals their emissivity. Explicitly, if a cloud emitting thermal radiation with a power per unit frequency equal to $\epsilon(\freq)\,u(\freq,\te)$, where $\epsilon(\freq)\in[0,1]$ is the emissivity at the frequency $\freq$ and $u(\freq,\te)$ is the power per unit frequency of a black body, then the absorbed CMB power per unit frequency equals $\epsilon(\freq)\,u(\freq,\tcmb)$. The effective CO-line luminosity measured against the CMB is the difference between the intrinsic luminosity emitted by the source and the luminosity absorbed from the CMB. If the intrinsic luminosity of the source is given by Eq.~(\ref{eqmodel1}), the effective luminosity against the CMB can be obtained by replacing the second factor (i.e.~the black body factor) in Eq.~(\ref{eqmodel1}) by
\be\label{eqell}
  \ell(J,\te,z)\!\equiv\!\frac{J^4}{\exp\!\left(\!\frac{\hp\fco J}{\kb\,\te}\!\right)\!-\!1}-\frac{J^4}{\exp\!\left(\!\frac{\hp\,\fco J}{\kb\tcmb(z)}\!\right)\!-\!1}.
\ee
In particular, this expression ensures that no radiation can be detected from molecular gas in thermal equilibrium with the CMB, since $\ell(J,\tcmb(z),z)=0$. Alternatively, Eq.~\ref{eqell} could also be expressed in terms of brightness temperatures \citep[e.g.~Eq.~(14.46) of][]{Rohlfs2004}.

\section{Combined model for CO-line luminosities}\label{subsection_flux}

A priori, we departed from the idea that CO-line fluxes $L_{\rm J}$ scale with the total mass of molecular hydrogen $\mhm$. Combining this assumption with the models of Sections \ref{subsection_cosed}--\ref{subsection_cmb_absorption}, we heuristically suggest that the CO-line luminosity (power) of the transitions $\transj$ is given by
\be\label{eqheuristiccomodel}
  L_{\rm J}=\mhm\cdot\frac{k\,Z\,\feff}{\overlap}\cdot\left[1-e^{-\overlap\,\optj}\right]\cdot\ell(J,\te,z),
\ee
where $k$ is an overall normalization factor. Consistent with common practice, we shall define $k$ in such a way that $L_{\rm J}$ is the line-power, obtained by integrating the power per unit solid angle emitted along the line-of-sight over all directions, even though the emission need not be isotropic. In fact, our model is explicitly non-isotropic, since the parameter $\overlap$ depends on the inclination of the galaxy via the axes-ratio $\qhm$.

To estimate the value of $k$, we note that in local regular galaxies all molecular gas is in the clumpy phase (i.e.~$\feff=1$), molecular clumps barely overlap (i.e.~$\overlap=1$), and the excitation temperatures are given by $\te\approx T_0$, hence $\optj=\optc=2$ and $\ell(1,\te,0)=2.45$. Therefore, Eq.~(\ref{eqheuristiccomodel}) for $J=1$ reduces to
\be\label{eqheuristicsimple}
  L_1 = 2.12\,k\,Z\,\mhm.
\ee
According to Eq.~(\ref{eqheuristicsimple}), $k$ is proportional to the standard CO/\hm~conversion factor $\alpha_1$ (Appendix \ref{appendix_backgroundx}); in fact, combining Eq.~(\ref{eqheuristicsimple}) with Eqs.~(\ref{eqdefalphaj}, \ref{eq4}), yields $k=11.9\,\kb\,fco^3\,Z^{-1}\,c^{-3}\,\alpha_1^{-3}$. Adopting the value $\alpha_1=4.6\,\msun({\rm K\,km\,s^{-1}\,pc^2})^{-1}$, typical for the MW \citep{Solomon2005}, and a cold gas metallicity of $Z=0.01-0.02$, yields $k=5-10\cdot10^{-8}\rm\,W\,kg^{-1}$.

\begin{figure*}[t!]
  \includegraphics[width=17cm]{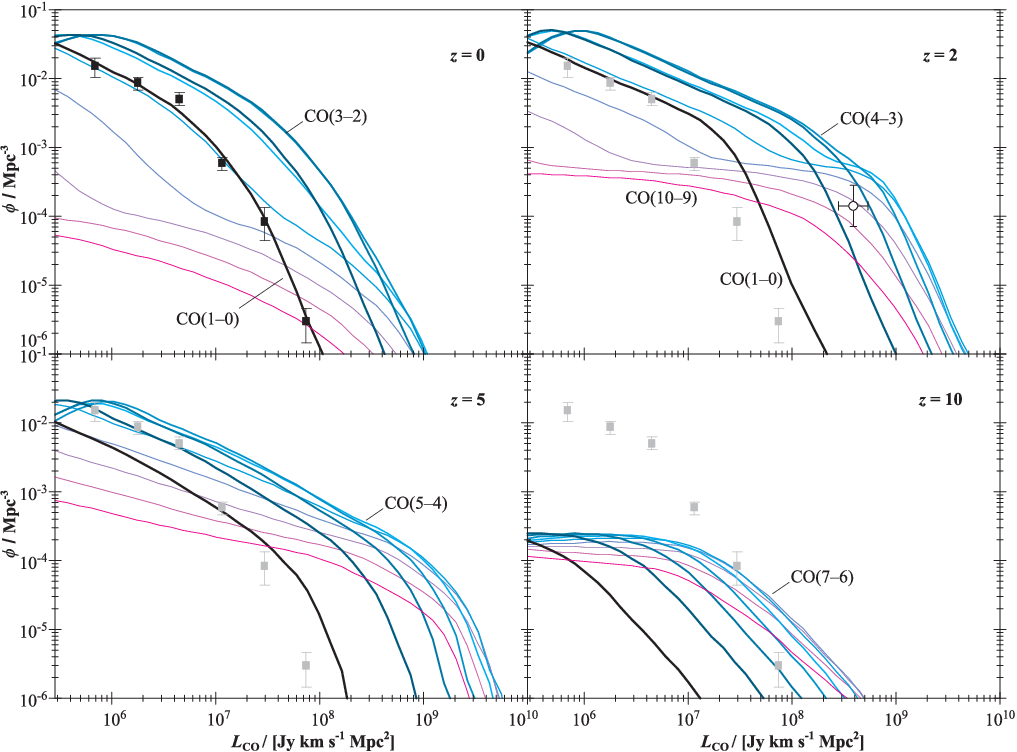}
  \caption{Predicted cosmic evolution of the CO-LFs in the redshift range $z=0-10$. The thick black line represents the CO(1--0) transition, while increasingly thin and red lines represent the increasingly higher order transitions up to CO(10--9). Filled points and error bars represent the observed CO(1--0)-LF of the local universe \citep{Keres2003}. The open circle with error bars corresponds to the CO(2--1) density estimate based on two detections in regular galaxies at $z\approx1.5$ by \cite{Daddi2008} (see \citealp{Obreschkow2009c} for further explanations). The differential space density $\phi(\lvel)$ is defined as the number of sources per unit comoving volume and unit $\log_{10}(\lvel)$ with a velocity-integrated luminosity $\lvel$.}
  \label{fig_colf_evolution}
\end{figure*}

Here, we shall fix the parameter $k$ such that Eq.~(\ref{eqheuristiccomodel}) applied to the \hm-masses of our simulated galaxies (Section \ref{section_simulation}) at $z=0$ reproduces the observed CO(1--0)-LF of the local universe as measured by \cite{Keres2003} (see Fig.~\ref{fig_colf_evolution}). A $\chi^2$-minimization for the luminosity range $L_1>10^{5.5}\rm\,Jy\,km\,s^{-1}\,Mpc^2$, i.e.~the range where our simulated \hm-MF is complete, yields
\be\label{eqkconst}
  k = 8\cdot10^{-8}\rm\,W\,kg^{-1},
\ee
which is indeed consistent with the aforementioned value predicted from the CO/\hm~conversion of the MW.

\section{Results}\label{section_results}

We have applied the model of Eq.~(\ref{eqheuristiccomodel}) to the galaxies of the hydrogen simulation described in Section \ref{section_simulation}. The predicted CO-LFs for the first 10 rotational transitions in the redshift range $z=0-10$ are displayed in Fig.~\ref{fig_colf_evolution}. For consistency with observer's practice, the luminosity scales refer to velocity-integrated luminosities $\lvel$, as opposed to the frequency-integrated luminosities $L$ (= power) used in Section \ref{section_method}. The conversion between those luminosities depends on the wavelength of the emission line as explained in Appendix \ref{appendix_luminosities}.

The good match (reduced $\chi^2=0.7$) between the simulated CO(1--0)-LF at $z=0$ and the local CO(1--0)-LF, inferred by \cite{Keres2003} from FIR-selected sample of IRAS galaxies, is due to our tuning of the constant $k$ and the reasonably accurate \hm-mass distribution of our hydrogen simulation (Section \ref{section_simulation}).

The simulation shows a clear signature of cosmic downsizing from $z=2$ to $z=0$ for all CO-transitions. This feature reflects the predicted downsizing of \hm-masses \citep{Obreschkow2009c}. For the particular case of the CO(2--1)-LF at $z=2$, the simulation result is roughly consistent with the space density (open circle in Fig.~\ref{fig_colf_evolution}) inferred from two recent CO(2--1)-emission measurements in normal galaxies at $z=1.5$ by \cite{Daddi2008}. We note that the offset of this empirical data point from our simulation is larger in Fig.~\ref{fig_colf_evolution} than in Fig.~1 of \cite{Obreschkow2009c}. In the latter, we have compared the empirical data point of \cite{Daddi2008} to our simulated \hm-MF at $z=2$. To this end we converted the measured CO(2--1)-luminosities into \hm-masses using the standard CO/\hm~conversion factor for ULIRGs, $\alpha_2=1~\msun(\rm K~km~s^{-1}\,pc^{-2})^{-1}$ (\citealp{Daddi2008}; see definition of $\alpha_2$ in Appendix \ref{appendix_backgroundx}). However, the model for the CO/\hm~conversion of this paper yields higher values of $\alpha_2$ for regular high-redshift galaxies, such as those found by \cite{Daddi2008}.

For higher order transitions ($J>5$), the predicted downsizing even extends out to $z\approx5$, due to the strong dependence of these transitions on SBs and AGNs (Section \ref{subsection_effect_sbag}). In Fig.~\ref{fig_colf_evolution}, the ``dominant'' transition, i.e.~the one with the maximal velocity-integrated luminosity per unit cosmic volume, is indicated at each redshift. The upper $J$-level of this transition increases with redshift due to the combined radiative heating by SBs and AGNs. We shall now analyze the dependence of the CO-LFs on the individual mechanisms modeled in Section \ref{section_method}.

\subsection{Effects of radiative heating by SBs and AGNs}\label{subsection_effect_sbag}

\begin{figure}[t]
  \includegraphics[width=\columnwidth]{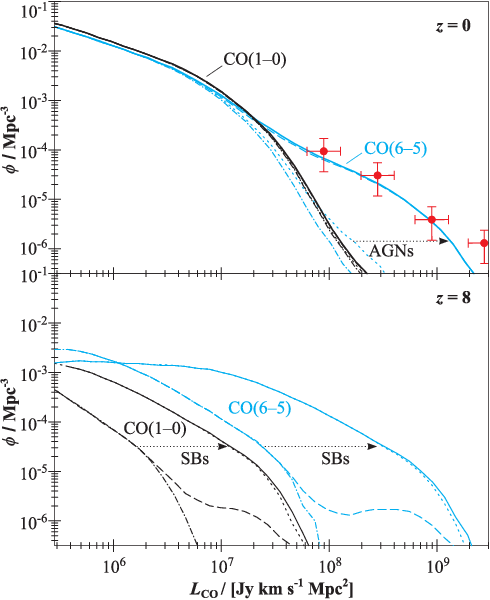}
  \caption{Effects of SB- and AGN-heating on the CO(1--0)-LF (black) and CO(6--5)-LF (blue) at redshifts $z=0$ and $z=8$. The solid lines represent the CO-LFs of the full model, such as shown in Fig.~\ref{fig_colf_evolution}, while the other lines represent the cases where either SBs (dashed), AGNs (dotted), or both (dash-dotted) were suppressed in the simulation. The red dots with error bars represent the local HX-LF \citep{Oshima2001}, mapped onto the CO(6--5)-luminosity scale as explained in Section \ref{subsection_effect_sbag}.}
  \label{fig_effect_sbagn}
\end{figure}

Fig.~\ref{fig_effect_sbagn} compares the simulated CO(1--0)-LF and CO(6--5)-LF at $z=0$ and $z=8$ to the corresponding LFs, if either AGN-heating or SB-heating is suppressed. At low redshift, both SBs and AGNs have nearly no observable effect on the CO(1--0)-LF, consistent with the conclusion of \cite{Keres2003} that only the highest luminosity-bin of the measured CO(1--0)-LF could indicate a deviation from a Schechter-function distribution, perhaps due to SBs in the sample.

By contrast, the CO(6--5)-LF appears to be significantly boosted by AGNs at $z=0$. In fact, this simulated LF deviates from a Schechter function and exhibits two ``knees'', respectively corresponding to a ``normal'' galaxy population (left knee) and a more luminous population heated by AGNs (right knee). Since the luminous end of the CO(6--5)-LF is entirely dominated by AGN-heating, we expect the local space density of the most CO(6--5)-luminous objects to match the space density of local AGNs. To test the simulation, we therefore overlaid the simulated local CO(6--5)-LF with the most recent empirical determination of the local hard ($2-8\rm\,keV$) X-ray-LF (HX-LF) obtained by \cite{Yencho2009} (data points in Fig.~\ref{fig_effect_sbagn}, top). This HX-LF relies on a galaxy sample studied by the X-ray \emph{Chandra} observatory. In order to map the HX-luminosity scale onto the CO(6--5)-luminosity scale, we crudely assumed a proportional relation between the two, tuned to the empirical data from the Cloverleaf quasar. We evaluated the lensed HX-luminosity (at $2-8\rm\,keV$ rest-frame) of the Cloverleaf quasar directly from the X-ray SED measured and corrected for Galactic absorption by \cite{Oshima2001}. For the cosmology of this paper, this HX-luminosity is $L_{\rm HX}=(1\pm0.5)\cdot10^{45}\rm\,erg\,s^{-1}$. On the other hand, the lensed CO(6--5)-line luminosity of the Cloverleaf quasar, interpolated from the CO(5--4) and CO(7--6) line fluxes presented by \cite{Barvainis1997} and corrected for the standard cosmology of this paper, amounts to $L_{\rm J=6}\approx5\cdot10^{10}\rm\,Jy\,km\,s^{-1}\,Mpc^2$. Assuming that the $L_{\rm HX}/L_{\rm J=6}$-ratio of the Cloverleaf quasar is not affected by differential magnification and that it mimics the $L_{\rm HX}/L_{\rm J=6}$-ratio of local AGNs, the HX-LF \citep{Yencho2009} transforms into the data points shown in Fig.~\ref{fig_effect_sbagn}. The vertical error bars represent the statistical density uncertainties given for the HX-LF, while the horizontal error bars represent the $50\%$ uncertainty of $L_{\rm HX}$. The good fit between the space densities of local AGNs and those predicted for the luminous CO(6--5)-sources supports our prediction.

At very high redshift ($z\gtrsim7$), where the predicted space density of AGNs in the DeLucia-catalog is extremely low \citep[see][]{Croton2006}, SBs become the dominant source of CO-heating as shown in Fig.~\ref{fig_effect_sbagn} (bottom). This analysis predicts that SB-heated molecular gas disks are the most likely objects to be detected in CO-line emission at $z\gtrsim7$. The optimal transitions are CO(8--7) and CO(6--5) in terms of velocity-integrated luminosities or surface-brightness temperatures, respectively.

The effects of gas heating by the CMB will are discussed together with the effects of the CMB as an observing background in Section \ref{subsection_effects_cmb}.

\subsection{Effects of overlapping molecular gas}\label{subsection_effects_overlap}

\begin{figure}[t]
  \includegraphics[width=\columnwidth]{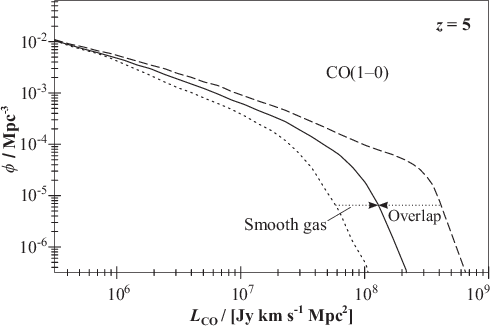}
  \caption{Effects of overlapping clumps and smoothly distributed gas on the CO(1--0)-LF at $z=5$. The solid line represents the CO-LF of the full model, such as shown in Fig.~\ref{fig_colf_evolution}. The other lines correspond to the suppression of overlap effects (dashed) and smoothly distributed gas (dotted). Note the different scale of the axes compared to the other figures.}
  \label{fig_effect_density}
\end{figure}

The effect of overlapping clumps (Section \ref{subsection_overlap}) exhibits a modest dependence on the upper $J$-level of the CO-transitions, although minor differences may occur due to the dependence of the optical depths on $J$ (see Eq.~\ref{eqmodel2}). Our model predicts that the effect of overlapping clouds becomes increasingly important with redshift, as a direct consequence of the predicted increase in the surface densities of galaxies with redshift. Between $z=0$ and $z=1$, the effect is negligible (i.e.~$<0.1\rm\,dex$ luminosity change), while at $z=5$ CO-luminosities are predicted to be reduced by a factor 2--3 due to cloud overlap. The dashed line in Fig.~\ref{fig_effect_density} illustrates the effect of ignoring the overlap of clumps (by forcing $\overlap=1$) at $z=5$.

\subsection{Effects of smooth molecular gas}\label{subsection_effects_smooth}

In our model (see Section \ref{subsection_densegas}), we assumed that molecular gas in very dense galaxy-parts is smoothly distributed, rather than organized in gravitationally bound GMCs. Within our simplistic treatment (Eqs.~\ref{eqmeff}, \ref{eqheuristiccomodel}), this effect is independent of the $J$-level of the CO-transition. The effect of smooth gas becomes increasingly important with redshift, as a direct consequence of the predicted increase in the surface densities of galaxies with redshift. Between $z=0$ and $z=1$ the effect is negligible (i.e.~$<0.1\rm\,dex$ luminosity change), but at $z=5$ its importance is comparable to that of heating by strong SBs and massive AGNs. The dotted line in Fig.~\ref{fig_effect_density} shows the effect of ignoring the possibility of smooth gas (by forcing $\fclumpy=1$) at $z=5$.

\subsection{Effects of metallicity}

\begin{figure}[t]
  \includegraphics[width=\columnwidth]{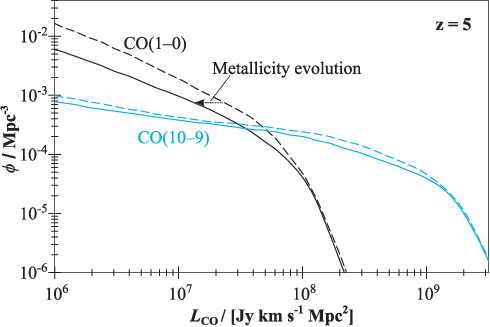}
  \caption{Effects of the cosmic evolution of cold gas metallicity on the LFs of CO(1--0) (black) and CO(10--9) (blue) at $z=5$. Solid lines represent the CO-LFs of the full model, such as shown in Fig.~\ref{fig_colf_evolution}, while dashed lines represent the CO-LFs, where the cosmic evolution of metals has been suppressed in the simulation.}
  \label{fig_effect_metals}
\end{figure}

Fig.~\ref{fig_effect_metals} shows the effect of neglecting the cosmic evolution of cold metals in galaxies, by illustrating the effects at $z=5$ of suppressing this evolution. In general, the effect of metallicity appears to be relatively weak, since the cosmic evolution of the cold gas metallicity from $z=5$ to $z=0$ is relatively weak as discussed in more detail in Section 6.3 of \cite{Obreschkow2009b}.

We also note that the cosmic evolution of the cold gas metallicity has a stronger effect on weak CO-sources than on the luminous ones. In fact, galaxies more luminous than the ``knee'' of the CO-LFs are nearly unaffected by the cosmic evolution of metals at $z=5$ compared to $z=0$. The reason for this feature is that the more CO-luminous galaxies are, on average, more massive and older, and hence they have already formed the bulk of their metals at $z>5$.

\subsection{Effects of the CMB}\label{subsection_effects_cmb}

\begin{figure}[t]
  \includegraphics[width=\columnwidth]{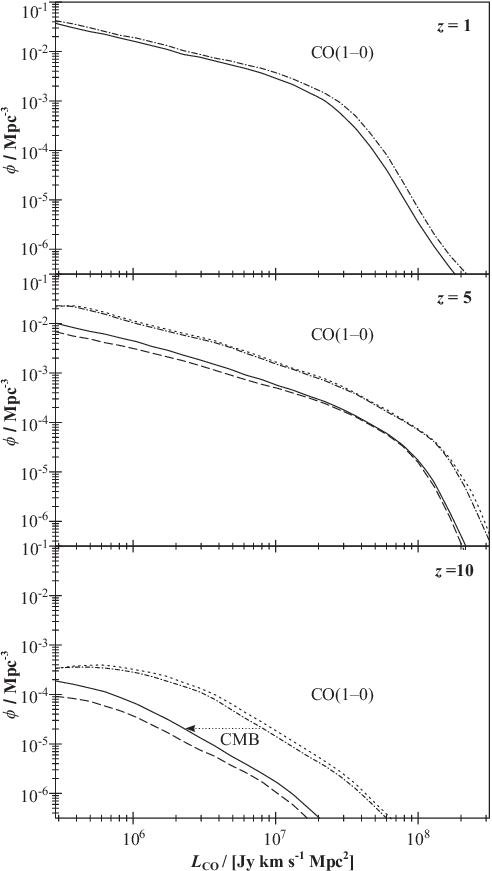}
  \caption{Effects of the CMB on the CO(1--0)-LF at $z=1$, $z=5$, and $z=10$. The solid lines represent the CO-LFs corresponding to the full model, such as shown in Fig.~\ref{fig_colf_evolution}. The other lines represent the cases, where either the gas-heating by the CMB (dashed), the CMB as an observing background (dotted), or both effects (dash-dotted) have been suppressed in the simulation.}
  \label{fig_effect_cmb}
\end{figure}

The CMB as an observing background already becomes noticeable at $z\approx1$, where its effective reduction of the CO(1--0)-luminosities amounts to about $0.1\rm\,dex$ (less for higher order transition) according to Eq.~(\ref{eqell}). This effect increases steeply with redshift and reaches $1\rm\,dex$ at $z\approx5$ for the CO(1--0) line, such as shown in Fig.~\ref{fig_effect_cmb}. The increase of the CO-excitation temperature by the heating effect of the CMB appears to be a minor effect, which only becomes noticeable around $z\approx5$. This effect acts against the loss of detectable luminosity by the CMB as an observing background by about $10\%$ at $z\approx5$ for the CO(1--0) line (slightly more for higher order transitions).

Our model generally predicts that the effect of the CMB as an observing background always dominates the opposite effect of the CMB as a source of heating. Hence, the combined effect of the CMB always reduces the detectable luminosities of CO-sources, at all redshifts and for all rotational transitions. This effect is most pronounced for lower order transitions, since emission from higher order transitions originates mostly from SBs and AGNs, whose heating effect can mask the comparatively low temperature of the CMB (e.g.~$\tcmb\ll\tsb$ and $\tcmb\ll\tagn$). For regular galaxies (no SBs, no AGNs), the combined effect of the CMB increases with redshift in such a way that these galaxies become virtually invisible in CO-line emission against the CMB at $z\gtrsim7$.

This result contradicts the claims of \cite{Silk1997} and \cite{Gnedin2001} that the higher excitation temperatures caused by the warm CMB of the early universe will ease the detection of CO-emission lines. The conclusion of these authors only accounts for gas heating by the CMB, but it ignores the CMB as an inevitable observing background. The importance of the CMB as an observing background has already been emphasized by \cite{Combes1999} and \cite{Papadopoulos2000}.

\vfill
\section{Discussion}\label{section_discussion}

\subsection{Ranking of various mechanisms}

The analysis of Section \ref{section_results} can be summarized in a ranking list of the different mechanisms affecting the CO/\hm~conversion. This ranking naturally depends on the redshift, the considered part of the CO-LF, and the $J$-level of the CO-transition. Here we consider the source population around the ``knee'' of the CO-LFs for the levels $J=3-6$ at redshift $z\approx3$. This case corresponds to using the fully funded \citep{Maiolino2008} ALMA-bands 3 and 4 for the first ALMA-science goal, i.e.~the detection of a MW-type galaxy in CO-line emission at $z\approx3$ \citep{DeBreuck2005}. For this particular setting the predicted ranking (from most important to least important) is
\begin{enumerate}
  \item Gas heating by AGNs $(+)$
  \item Gas heating by SBs $(+)$
  \item Overlap of clumps $(-)$
  \item Smooth gas $(+)$
  \item CMB as an observing background $(-)$
  \item Cosmic evolution of the cold gas metallicity $(-)$
  \item Gas heating by the CMB $(+)$
\end{enumerate}
The signs in parentheses indicate whether the effect increases $(+)$ or decreases $(-)$ the CO-line luminosities. Although this ranking may change considerably with redshift and with $J$ (e.g.~Fig.~\ref{fig_effect_sbagn}), the above ranking can be considered as a rule of thumb for estimating the relative importance of various effects. For example, if a simulation of CO-LFs includes a model for the smooth gas in high-redshift galaxies, then it should also account for the heating by SBs and AGNs and self-shielding by overlapping clumps.

\subsection{Model limitations}\label{subsection_limitations}

The predictions presented is this paper are approximate ramifications of a semi-empirical model, which potentially suffers from simplifications and uncertainties on each of the four successive simulation-layers: (i) the Millennium dark matter simulation, (ii) the semi-analytic galaxy simulation of the DeLucia-catalog, (iii) our post-processing to assign extended \ha- and \hm-properties to each galaxy, (iv) the model for CO-line emission introduced in this paper. It is beyond the scope of this paper to discuss the limitations related to the simulation-layers (i)--(iii), but extensive discussions were provided by \cite{Springel2005}, \cite{Croton2006}, and \cite{Obreschkow2009b}, respectively.

All four simulation-layers were widely constrained by a broad variety of observations: (i) the cosmological parameters for the Millennium simulation were adopted from 2dFGRS \citep{Colless2001} and WMAP \citep{Spergel2003,Bennett2003}; (ii) the semi-analytic recipes are motivated by various references given in \cite{Croton2006} and the free parameters were tuned to fit the luminosity/colour/morphology distribution of low-redshift galaxies \citep{Cole2001,Huang2003,Norberg2002}, the bulge-to-black hole mass relation \citep{Haering2004}, the Tully--Fisher relation \citep{Giovanelli1997}, the cold gas metallicity as a function of stellar mass \citep{Tremonti2004}; (iii) our model to assign \ha- and \hm-properties is motivated by various studies of \ha~and \hm~in local galaxies \cite[e.g.][]{Young2002,Blitz2006,Leroy2008,Elmegreen1993} and the free parameter was tuned to the local space density of cold gas \citep{Keres2003,Zwaan2005,Obreschkow2009a}; (iv) our CO-model was constrained as described in Section \ref{section_method} of this paper. Since this empirical basis is widely dominated by observations in the local universe, we expect our low-redshift predictions for CO, such as the CO-LFs for higher order transitions, to be more accurate than the high-redshift predictions.

With regard to our model for CO-line emission, the most reliably modeled effects are those of the CMB, since they could be assessed from global considerations, with no strong dependence on free parameters (see Sections \ref{subsection_cmb_absorption}, \ref{subsection_effect_sbag}). Also the effects of cold gas metallicity are relatively unproblematic: first, these effects are relatively small (e.g.~Fig.~\ref{fig_effect_metals}); second, the correlation between metallicity and the CO/\hm~conversion is empirically supported (see Section \ref{subsection_metallicity}); third, the metallicities predicted by the DeLucia-catalog seem reliable as they reproduce the mass--metallicity relation inferred from 53,000 star forming galaxies in the Sloan Digital Sky Survey \citep{DeLucia2004,Croton2006,Tremonti2004}.

The effects of inter-cloud heating by high density star formation (or SBs) and heating by AGNs are much less certain. Our temperature model relies on the CO-SEDs of only 7 galaxies (4 SBs and 3 QSOs) with poorly constrained star formation densities and black hole accretion rates. It is further possible that the molecular material in some of these galaxies is simultaneously heated by both a SB and an AGN. The relations of Eqs.~(\ref{eqexcitationt}--\ref{eqparatagn}) are simplistic parameterizations of our limited knowledge on gas heating by SBs and AGNs, but they may require a revision as larger galaxy samples with simultaneous CO-SEDs, SFRs, and black hole accretions rates come on line. Given the present-day uncertainties of SB- and AGN-heating, the use of the LTE-model for CO-SEDs (Section \ref{subsection_effect_sbag}) seems sufficient. In fact, the offset of the LTE-model model from the more complex LVG-models is small \citep{Combes1999} compared to the systematic uncertainties of radiative heating by SBs and AGNs.

Surprisingly, we found that self-shielding by overlapping clumps is perhaps the most subtle effect to model at $z>3$, because it seems to be a very significant effect (see Fig.~\ref{fig_effect_density}) and yet its physical complexity is considerable. Especially in the case of galaxies with heavily overlapping clumps (i.e.~$\overlap\gg1$), the value of the overlap parameter $\overlap$ sensibly depends on the radius and mass of molecular clumps. If we also consider that clumps are not randomly distributed, but organized in cloud-complexes, and that their geometries are far from spherical, the predicted CO-line luminosities of galaxies with heavily overlapping clumps could differ from our current prediction by nearly an order of magnitude. Similar uncertainties should be assumed for the effects of smooth gas in high-redshift galaxies. In fact, the critical surface density $\sigcrit$, at which gas transforms from clouds to smooth disks, is very uncertain and may vary as a function of the mass and size of the galaxy.

\section{Conclusion}\label{section_conclusion}

We have predicted the cosmic evolution of the galaxy LFs for the first 10 rotational transitions of the CO-molecule. This prediction relies on a combination of a recently presented simulation of \hm-masses in $\sim3\cdot10^7$ evolving galaxies with a model for the conversion between \hm-masses and CO-line luminosities. The latter model accounts for radiative heating by AGNs, SBs, and the CMB, for smooth and overlapping gas, for the cosmic evolution of metallicity, and for the CMB as an observing background.

The main outcome of this study is two-fold. Firstly, the predicted CO-LFs are probably the most robust basis to-date towards predicting the CO-line detections of high-redshift surveys with future telescopes, such as ALMA \citep[see][]{Blain2000}, the LMT, or phase-3 of the Square Kilometre Array (SKA). Secondly, this study revealed that the most serious uncertainties of the CO-LFs at high redshifts originate from the poorly understood self-shielding of overlapping clouds, from the smooth gas in luminous galaxies, and from the heating by SBs and AGNs. Hence, any serious progress in predicting the CO-LFs must address these mechanisms in more detail. By contrast, the widely cited effects of the CMB and the cosmic evolution of metallicity seem to be relatively well modeled.

This study makes some explicit predictions, which could be tested in future CO-surveys; e.g.:
\begin{enumerate}
  \item The CO-LFs associated with the first 10 rotational transitions should show a strong signature of ``downsizing'' in the redshift range $z=0-2$. Explicitly, the total power of each CO-line per comoving volume increases from $z=0$ to $z=2$ by a factor 2 to more than 10, depending on the CO-transition (see Fig.~\ref{fig_colf_evolution}).
  \item On average, the relative CO-line power in higher order transitions, i.e.~the excitation temperature $\te$, increases monotonically with redshift $z$. This is a consequence of more heating at high $z$, mainly due to SBs and AGNs (see Fig.~\ref{fig_colf_evolution}).
  \item Some CO-LFs (e.g.~CO(6--5) and CO(7--6) at $z=0$, and CO(5--4), CO(6--5) and CO(7--6) at $z=2$) significantly deviate from a Schechter function. They are predicted to have two ``knees'', respectively corresponding to a ``normal'' galaxy population and a more CO-luminous population, where the gas is heated mostly by AGNs (e.g.~Fig.~\ref{fig_effect_sbagn} top).
  \item Out to the most distant galaxies, most of the cosmic CO-luminosity is predicted to stem from regular clumpy gas, i.e.~from GMCs, rather than a hypothetical dense phase, which is believed to dominate some ULIRGs \citep{Downes1993,Downes1998}.
  \item The CMB will significantly suppress the apparent CO-line flux of galaxies at high $z$ (see Fig.~\ref{fig_effect_cmb}). In particular, galaxies at $z\gtrsim7$ with no strong source of internal heating, such as a SB or an AGN, will not be detectable in CO-line emission.\vspace{1cm}
\end{enumerate}

\acknowledgments
This effort/activity is supported by the European Community Framework Programme 6, Square Kilometre Array Design Studies (SKADS), contract no 011938. The Millennium Simulation databases and the web application providing online access to them were constructed as part of the activities of the German Astrophysical Virtual Observatory. We thank W.~F.~Wall and the anonymous referee for helpful suggestions.


\appendix

\section{A. Luminosities and fluxes of lines}\label{appendix_luminosities}

This section overviews the concepts used in relation with line fluxes and line luminosities with an emphasis on connecting the terminology and units of observers to those of theoreticians.

\subsection{Terminology and definitions}

Any continuous isotropic electromagnetic radiation from a point-source is completely characterized by the luminosity density (or monochromatic luminosity) $l(\nu)$, an intrinsic quantity measured in units proportional to ${\rm1\,W\,Hz^{-1}\equiv1\,J}$. The corresponding observable quantity is the flux density (or monochromatic flux) $s(\nu)$, measured in units proportional to ${\rm1\,W\,Hz^{-1}\,m^{-2}\equiv1\,kg\,s^{-2}}$ (${\rm 1\,Jy = 10^{-26}\,W\,Hz^{-1}\,m^{-2}}$).

The luminosity distance $\dl$ is defined in such a way, that the conservation of energy applied to $l(\nu)$ and $s(\nu)$ takes the standard form of the continuity equation,
\be\label{eqconservation}
  \int l(\nu){\rm d}\nu = 4\pi\dl^2\int s(\nu){\rm d}\nu.
\ee

By definition a source is at redshift $z$, if electromagnetic radiation emitted by this source at a rest-frame frequency $\frest$ is observed at a frequency $\fobs=\frest\,(1+z)^{-1}$; or, in terms of wave-lengths, $\wlobs=\wlrest\,(1+z)$. A frequency interval ${\rm d}\nu$ around $\frest$ will be compressed to ${\rm d}\nu(1+z)^{-1}$, when observed at $\fobs$; therefore,
\be\label{eqdensities}
  l(\frest) = 4\pi\dl^2 s(\fobs)\,(1+z)^{-1}.
\ee

If the source presents an emission line centered at a rest-frame frequency $\frest$, one often considers the integrated luminosity and flux from the whole line. However, at least three definitions of these integrated quantities are commonly used. The most physically meaningful choices are the frequency-integrated quantities,
\be\label{eqdeffreqint}
  \lfreq\equiv\int_{\frest-\Delta\frest}^{\frest+\Delta\frest}\!\!\!\!l(\nu){\rm d}\nu,~~~\sfreq\equiv\int_{\fobs-\Delta\fobs}^{\fobs+\Delta\fobs}\!\!\!\!s(\nu){\rm d}\nu,
\ee
where $\Delta\frest$ and $\Delta\fobs$ represent the half-widths of the line in rest-frame frequency and observer-frame frequency, respectively. The precise definition of these half-widths (i.e.~the definition of where the line ends) depends on the observer's choice. $\lfreq$ represents the actual power of the emission line and is measured in units proportional to $\rm 1\,W$ ($1\,\lsun=3.839\cdot10^{26}\,\rm W$). $\sfreq$ represents the power per unit area received by the observer, measured in units proportional to ${\rm 1\,W\,m^{-2}\equiv1\,kg\,s^{-3}}$.

An alternative definition to the frequency-integrated quantities are the velocity-integrated analogues, often preferred by observers,
\be\label{eqdefvelint}
  \lvel\equiv\int_{-\Delta\vel}^{+\Delta\vel}\!\!\!\!l(\nu){\rm d}\vel,~~~\svel\equiv\int_{-\Delta\vel}^{+\Delta\vel}\!\!\!\!s(\nu){\rm d}\vel,
\ee
where $V$ is the rest-frame velocity, projected on the line-of-sight, of the emitting material relative to the center of the observed galaxy, and $\Delta\vel$ is the maximal velocity (rotation+dispersion) of the emitting material. $\lvel$ is measured in units proportional to $\rm 1\,kg\,m^3\,s^{-3}$ ($\rm1\,Jy\,km\,s^{-1}\,Mpc^2=9.521\cdot10^{21}\,kg\,m^3\,s^{-3}$), and $\svel$ is measured in units proportional to $\rm 1\,kg\,m\,s^{-3}$ ($\rm 1\,Jy\,km\,s^{-1}=10^{-23}\,kg\,m\,s^{-3}$).

\begin{figure}[h]
\begin{center}
  \includegraphics[width=7cm]{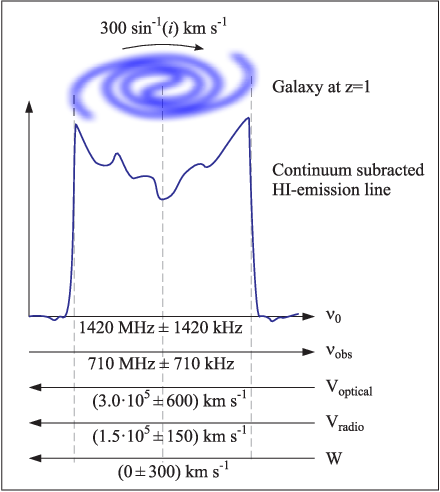}
\end{center}
  \caption{Different velocity measures used in relation to emission lines from galaxies.}
  \label{fig_velocity_definitions}
\end{figure}

Confusion sometimes arises in the definition of the velocity $\vel$ in Eqs.~(\ref{eqdefvelint}), since several definitions of velocity are commonly used in the context of emission and absorption lines (see Fig.~\ref{fig_velocity_definitions}): (i) the standard recession velocity $\vel_{\rm opt}(\freq)=c(\frest-\freq)/\freq$, traditionally used by optical astronomers; (ii) the variation $\vel_{\rm radio}(\freq)=c(\frest-\freq)/\frest$, sometimes employed by radio-astronomers; (iii) the ``intrinsic rest-frame velocity'' $\velint$, representing the rest-frame velocity, projected on the line-of-sight, of the emitting material relative to the center of the observed galaxy. For the investigation of emission (or absorption) lines at high redshift, it is critical to specify, which definition of the velocity $\vel$ is used in the definition of velocity-integrated quantities like $\lvel$ and $\svel$. The most natural choice, which we adopted in Eqs.~(\ref{eqdefvelint}), is $\vel=\velint$. This is the only choice, which makes $\lvel$ an intrinsic property, that does not depend on the observer's distance.

In the rest-frame of the observed galaxy, the center of the emission line is at the frequency $\frest$ and $\vel$ is computed as $\vel=c(\freq-\frest)/\frest$. Hence,
\be\label{eqjacobianrest}
  \frac{{\rm d}\vel}{{\rm d}\freq} = \wlrest~~{\rm if}~\freq~{\rm \text{is in the rest-frame}}.
\ee
In the observer's frame, the center of the emission line is at the frequency $\fobs$ and and $\vel$ is computed as $\vel=c(\freq-\fobs)/\fobs$. Hence,
\be\label{eqjacobianobs}
  \frac{{\rm d}\vel}{{\rm d}\freq} = \wlobs~~{\rm if}~\freq~{\rm \text{is in the observer-frame}}.
\ee

Sometimes, line luminosities are defined with respect to the surface brightness temperature $\tb$, which is defined as the (frequency-dependent) temperature of a black-body with the physical size of the observed source and providing an identical flux density $s(\fobs)$. In radio astronomy the energy distribution of a black body is commonly approximated by the Rayleigh-Jeans law, i.e.~the power radiated per unit of surface area, frequency, and solid angle equals $u(\freq)=2\freq^2 k\tb c^{-2}$, where $\kb$ is the Boltzmann constant. For isotropic line emission at rest-frame frequency $\frest$ this implies $l(\frest)=4\pi\da^2\Omega\,u(\frest)=8\pi\frest^2\kb\tb c^{-2}\da^2\Omega$, where $\da=\dl(1+z)^{-2}$ is the angular diameter distance and $\Omega$ is the solid angle subtended by the source. Using Eq.~(\ref{eqdensities}), we then obtain
\be\label{eqtb}
  \tb(\fobs) = \frac{c^2}{2\,\kb}\,\frac{s(\fobs)\,\frest^{-2}\,(1+z)^3}{\Omega}.
\ee
$\tb$ is an intrinsic quantity, which does not change with redshift $z$, as can be seen from $s(\fobs)/\Omega\propto(1+z)^{-3}$. Often the brightness temperature intensity $\ibt$ of an emission line source is defined as the velocity-integrated brightness temperature,
\be\label{eqdefibt}
  \ibt\equiv\int_{-\Delta\vel}^{+\Delta\vel}\!\!\!\!\tb(\freq){\rm d}\vel,
\ee
giving units proportional to $\rm 1\,K\,m\,s^{-1}$ ($\rm 1\,K\,km\,s^{-1}=10^3\,K\,m\,s^{-1}$). Alternatively, observers sometimes define the intensity $\ibt$ as the velocity-integral of the ``beam-diluted'' brightness temperature $T_{\rm mb}$, which is smaller than $\tb$ if the source does not cover the whole beam. We also note that some authors use the symbol $I$ for fluxes, which we label $S$ \citep[e.g.][]{Weiss2007}. The brightness temperature luminosity $\lbt$ is defined as the product of the intensity and the source area \citep{Solomon1997},
\be\label{eqdeflbt}
  \lbt \equiv \da^2\Omega\int_{-\Delta\vel}^{+\Delta\vel}\!\!\!\!\tb(\freq){\rm d}\vel.
\ee
This implies that $\lbt$ is measured in units proportional to $\rm 1\,K\,m^3\,s^{-1}$ ($\rm 1\,K\,km\,s^{-1}\,pc^2=9.521\cdot10^{35}\,K\,m^3\,s^{-1}$).

\subsection{Basic relations}\label{subsection_basic_relations}

From the conservation law of Eq.~(\ref{eqconservation}), we directly find the flux-to-luminosity relations
\be\label{eq1}
  \lfreq = 4\pi\dl^2\,\sfreq.
\ee

The velocity-integrated flux $\svel$ can be expressed in terms of the frequency-integrated flux $\sfreq$ by using the Jacobian of Eq.~(\ref{eqjacobianobs}),
\be\label{eq2}
  \svel = \wlobs\,\sfreq.
\ee
Similarly, the velocity-integrated luminosities $\lvel$ can be expressed in terms of the frequency-integrated luminosity $\lfreq$ (i.e.~intrinsic power of the emission line) by using the Jacobian of Eq.~(\ref{eqjacobianrest}),
\be
  \lvel = \wlrest\,\lfreq, \label{eq3}\\
\ee
Finally, from Eqs.~(\ref{eqtb}, \ref{eqdeflbt}, \ref{eq1}, \ref{eq2}),
\be\label{eq4}
  \lbt = (8\pi\kb)^{-1}\wlrest^3\,\lfreq.
\ee

Using the four basic relations of Eqs.~(\ref{eq1}--\ref{eq4}), we can express any of the quantities $\lfreq$, $\sfreq$, $\lvel$, $\svel$, and $\lbt$ as a function of any other. For example, Eqs.~(\ref{eq1}, \ref{eq2}, \ref{eq3}) imply that
\be\label{eqfluxtolumvel}
  \lvel = (1+z)^{-1}\,4\pi\dl^2\,\svel.
\ee
Note that this relation differs from Eq.~(\ref{eq1}) by a redshift-factor. Eqs.~(\ref{eq1}, \ref{eq3}) imply that
\be\label{eqpresolomon1}
  \lfreq = \frac{4\pi}{c}\,\dl^2\fobs\svel,
\ee
or, using typical observer-units,
\be\label{eqsolomon1}
  \frac{\lfreq}{\lsun} = 1.040\cdot10^{-3}\,\left(\!\frac{\dl}{\rm Mpc}\!\right)^2\!\frac{\fobs}{\rm GHz}\,\frac{\svel}{\rm Jy\,km\,s^{-1}}\,.
\ee
This is equivalent to Eq.~(1) in \cite{Solomon2005}. Similarly, Eqs.~(\ref{eq1}, \ref{eq2}, \ref{eq4}) imply that
\be\label{eqpresolomon3}
  \lbt = \frac{c^2}{2\,\kb}\,\fobs^{-2}\,\dl^2(1+z)^{-3}\svel,
\ee
or, in observer units,
\be\label{eqsolomon3}
\begin{split}
  \frac{\lbt}{\rm K\,km\,s^{-1}\,pc^2} = 3.255\cdot10^7\left(\frac{\fobs}{\rm GHz}\right)^{-2}\left(\frac{\dl}{\rm Mpc}\right)^2\\
  \times\,(1+z)^{-3}\,\frac{\svel}{\rm Jy\,km\,s^{-1}}\,.\qquad\quad~
\end{split}
\ee
which is identical to Eq.~(3) in \cite{Solomon2005}.

\section{B. Background of the CO/\hm~conversion}\label{appendix_backgroundx}

To-date, most estimates of molecular gas masses in galaxies rely on radio and (sub-)millimeter emission lines of trace molecules, especially on emission lines associated with the decay of rotational excitations of the CO molecule.

It is not obvious that the CO-lines trace \hm, and this method has indeed a long history of controversy. From local observations in the MW, it has become obvious that molecular gas resides in loosely connected giant ($\sim10-100\rm\,pc$) ``clouds'', which are generally composed of hundreds of dense ``clumps'' ($\sim1\rm\,pc$), hosting even denser ``cores'' ($\sim0.1\rm\,pc$), where new stars are born (see e.g.~the Orion Molecular Cloud, \citealp{Maddalena1986,Tatematsu1993}). CO-line emission cannot be used as mass tracer of individual clumps and cores, since they are \emph{optically thick} to rotational CO-emission lines as can be inferred from the intensity-ratios between different rotational levels (\citealp{Binney1998} Chapter 8). However, if averaged over entire clouds or galaxies (typically $10^4-10^7$ clouds), CO behaves as if it were \emph{optically thin}, in a sense that individual clumps and cores do not significantly overlap (in space and frequency) \citep{Wall2006}, and hence on these large scales CO-line luminosities are expected to become suitable tracers of the molecular mass. Compelling empirical support for this conclusion was provided by the very tight correlation between the virial masses, estimated from sizes and velocity dispersions, and the CO(1--0)-luminosities of 273 molecular clouds in the MW analyzed by \cite{Solomon1987}.

To convert CO-line luminosities into \hm-masses, it is common to define the $X$-factors as
\be\label{eqdefxj}
  \xj\equiv\frac{\nhm}{\ibt_{\rm J}},
\ee
where $J$ is the upper rotational transition $\transj$, $\nhm$ is to column number-density of \hm-molecules (here, we exclude helium), and $\ibt_{\rm J}$ is the brightness temperature intensity [see definition in Eq.~(\ref{eqdefibt})] of the ${\rm CO}(\transj)$-emission line. Alternatively, the CO/\hm~conversion factors are sometimes defined as \citep{Solomon2005}
\be\label{eqdefalphaj}
  \alphaj\equiv\frac{\mhm}{\lbt_{\rm J}},
\ee
where $\lbt_{\rm J}$ is the brightness temperature luminosity [see definition in Eq.~(\ref{eqdeflbt})] of the ${\rm CO}(\transj)$-emission line. Note that the definitions of $\nhm$ and $\mhm$ in Eqs.~(\ref{eqdefxj}, \ref{eqdefalphaj}) do not include a helium fraction, but some authors \citep[e.g.][]{Downes1993} include a helium fraction of $\sim36\%$ in $\nhm$ and $\mhm$, which makes their values of $\xj$ and $\alphaj$ 1.36-times larger.

Since $\mhm=\da^2\Omega\nhm\masshm$, where $\masshm$ is the mass of a \hm-molecule, and $\lbt_{\rm J}=\da^2\Omega\ibt_{\rm J}$ for all $J\geq1$, we find that the two conversion factors are related by
\be\label{eqxalphaconversion1}
  \alphaj = \xj\,\masshm\,,
\ee
or, in typical observer units,
\be\label{eqxalphaconversion2}
  \frac{\alphaj}{\msun({\rm K\,km\,s^{-1}\,pc^2})^{-1}} = \frac{1.6\cdot10^{-20}\,\xj}{({\rm K\,km\,s^{-1}\,cm^2})^{-1}}.
\ee

From Eqs.~(\ref{eq3}, \ref{eq4}, \ref{eqdefalphaj}, \ref{eqxalphaconversion1}) it follows that
\be\label{eqmhmfromlvel1}
  \mhm = \frac{\masshm\,c^2}{8\pi\,k\,\freq_{\rm J}^2}\,\xj\,\lvel_{\rm J},
\ee
where $\lvel_{\rm J}$ and $\freq_{\rm J}$ respectively denote the velocity-integrated luminosity and the rest-frame frequency of the ${\rm CO}(\transj)$-emission line. $\freq_{\rm J}$ can be calculated as $\freq_{\rm J}=J\,\freq_{\rm CO}$ where $\freq_{\rm CO}=115\rm\,GHz$ is the rest-frame frequency of the CO(1--0)-line. Eq.~(\ref{eqmhmfromlvel1}) can then be expressed in typical observer units as
\be\label{eqmhmfromlvel2}
  \frac{\mhm}{\msun} = \frac{313\,J^{-2}\,\xj}{10^{20}({\rm K\,km\,s^{-1}\,cm^2})^{-1}}\cdot\frac{\lvel_{{\rm CO}(\transj)}}{\rm Jy\,km\,s^{-1}\,Mpc^2}.
\ee
Other mass--luminosity and mass--flux relations for \hm~commonly found in the standard literature can be derived from Eq.~(\ref{eqmhmfromlvel2}) and the basic relations in Section \ref{subsection_basic_relations}.

Both the theoretical and the empirical determination of this conversion have a long history in radio astronomy, and are still considered highly challenging problems at the present day (see overviews in~\citealp{Maloney1988}, \citealp{Wall2007}, \citealp{Dickman1986}). Different methods to measure $\alphaj$ (or $\xj$) were summarized by \cite{Downes1993}, \cite{Arimoto1996}, and \cite{Solomon2005}. The latter suggest that a sensible average value for the MW is $\alpha_1=3.4\msun({\rm K\,km\,s^{-1}\,pc^2})^{-1}$, or $\alpha_1=4.8\msun({\rm K\,km\,s^{-1}\,pc^2})^{-1}$, if helium is included in the definition of $\alpha_1$.

\section{C. Line emission of thermalized CO with finite optical depth}\label{appendix_thermal_model}

The rotational states of a diatomic molecule, such as CO, can be represented in the basis $\{\ket{J,m}\}$, where $J\geq0$ is the angular quantum number and $m\in\{-J,...,J\}$ is the magnetic quantum number. In the absence of external fields, the energy only depends on $J$ via $E_{\rm J}=\hp\,\fco\,J(J+1)/2$, where $\fco=115\rm\,GHz$ and is the rest-frame frequency of the transition $J=1\!\rightarrow\!0$. In local thermal equilibrium (LTE), the occupation probabilities of these energy-levels are therefore given by
\be
  n_{\rm J} = \frac{g_{\rm J}}{\partition(\te)}\,\exp\left[-\frac{\hp\,\fco\,J(J+1)}{2\,\kb\,\te}\right],
\ee
where $g_{\rm J}=2\,J+1$ are the degeneracies lifted by the quantum number $m$, $\te$ is the excitation temperature (which is identical to the kinetic gas temperature $T_{\rm k}$ in LTE-conditions), and $\partition(T)$ is the canonical partition function, which ensures the normalization condition $\sum{n_{\rm J}}=1$. The partition function is approximated to $<1\%$ for all $T>10\rm\,K$ by
\be
  \partition(\te) = \frac{2\,\kb\,\te}{\hp\,\fco}-\frac{2}{3}.
\ee

The interaction between a state $\ket{J,m}$ and the electromagnetic field only permits transitions simultaneously changing $J$ by $\pm1$ and $m$ by $-1,0,1$. To determine the electromagnetic emission emerging from the three transitions $\ket{J,m}\rightarrow\ket{J-1,m'}$, where $m'\in\{m,m\pm1\}$, we require a measure of the rates of spontaneous emission from $\ket{J,m}$, induced emission from $\ket{J,m}$, and absorption by $\ket{J-1,m'}$. These rates are effectively described by the Einstein coefficients $A_{\rm J,J-1}$, $B_{\rm J,J-1}$, and $B_{\rm J-1,J}$ (defined in \citealp{Binney1998}, Chapter 8), which can be calculated directly from the interaction Hamiltonian between the rotational states and the electromagnetic-field. From considerations of a gas in LTE it follows that these coefficients are related via $A_{\rm J,J-1}\propto J^3 B_{\rm J,J-1}$ and $g_{\rm J-1}B_{\rm J-1,J}=g_{\rm J}B_{\rm J,J-1}$. \cite{Rieger1974} showed that $A_{\rm J,J-1}$ scales with $J$ as
\be
  A_{\rm J,J-1} \propto \frac{J^4}{2\,J+1},
\ee
and hence
\be
  B_{\rm J,J-1}\propto\frac{J}{2\,J+1}\quad{\rm and}\quad B_{\rm J-1,J}\propto\frac{J}{2\,J-1}.
\ee

Following \cite{Binney1998}, the ``source function'' $l_{\rm J}$, which is proportional to the power radiated per unit frequency from the transition $\transj$ (i.e.~the sum of the power from all the transitions $\ket{J,m}\rightarrow\ket{J-1,m'}$) in an optically thick medium, is then given by
\be
  l_{\rm J} \propto \frac{n_{\rm J}\,A_{\rm J,J\!-\!1}}{n_{\rm J\!-\!1}\,B_{\rm J\!-\!1,J}\!-\!n_{\rm J}\,B_{\rm J,J\!-\!1}} \propto \frac{J^3}{\exp\left(\frac{\hp\,\fco\,J}{\kb\,\te}\right)-1},
\ee
and hence the \emph{frequency-integrated} power in a medium with arbitrary optical depth $\optj$ is given by
\be\label{eqlpropto}
\begin{split}
  L_{\rm J} \propto J\,l_{\rm J}\,\left[1-\exp(-\optj)\right] \\
  \propto\left[1-\exp(-\optj)\right]\cdot\frac{J^4}{\exp\left(\frac{\hp\,\freq_{\rm CO}\,J}{\kb\,\te}\right)-1},
\end{split}
\ee
where
\be\label{eqtaupropto}
\begin{split}
  \optj(\te)\propto J^{-1}\,(n_{\rm J\!-\!1}\,B_{\rm J\!-\!1,J}\!-\!n_{\rm J}\,B_{\rm J,J\!-\!1}) \\
  \propto \exp\!\left(-\frac{\hp\,\fco\,J^2}{2\,\kb\,\te}\right)\,\sinh\!\left(\frac{\hp\,\fco\,J}{2\,\kb\,\te}\right).
\end{split}
\ee

From Eq.~(\ref{eq4}) the brightness temperature luminosity is given by $L_{\rm J}^{\rm T}\propto L_{\rm J}J^{-3}$. In the particular case of an optically thick medium ($\optj\rightarrow\infty$) and high temperatures ($\kb\,\te\gg\hp\,\fco\,J$), Eq.~(\ref{eqlpropto}) then implies that $L_{\rm J}^{\rm T}$ is independent of $J$, which is indeed one of the essential properties of brightness temperature luminosities.

\end{document}